\documentclass[11pt]{article}

\usepackage{amsmath,amsfonts,bm}

\def\eqref#1{equation~\ref{#1}}

\def\1{\bm{1}}

\DeclareMathAlphabet{\mathsfit}{\encodingdefault}{\sfdefault}{m}{sl}
\SetMathAlphabet{\mathsfit}{bold}{\encodingdefault}{\sfdefault}{bx}{n}

\usepackage[]{acl}

\usepackage{times}
\usepackage{latexsym}
\usepackage[T1]{fontenc}
\usepackage[utf8]{inputenc}
\usepackage{microtype}

\usepackage{url}
\usepackage{graphicx}  
\usepackage{amsfonts}
\usepackage{CJK}
\usepackage{bm}
\usepackage{mathrsfs}
\usepackage{float}
\usepackage{xcolor,colortbl}
\usepackage{adjustbox}
\usepackage{subfigure}
\usepackage{booktabs,multirow}
\usepackage{bigstrut}
\usepackage{paralist}
\usepackage{bbm}
\usepackage{makecell}
\usepackage{nicefrac}       
\usepackage{microtype} 
\usepackage{epsfig}
\usepackage{diagbox}
\usepackage{framed}
\usepackage{empheq}
\usepackage{array}
\usepackage{enumitem}
\usepackage{mdwlist}
\usepackage{xspace,mfirstuc,tabulary}
\usepackage{tcolorbox}
\usepackage{algorithm}
\usepackage{placeins}
\usepackage{algpseudocode}
\usepackage{microtype}

\usepackage[normalem]{ulem}
\useunder{\uline}{\ul}{}
\usepackage{threeparttable}
\usepackage{makecell}
\usepackage{booktabs}

\usepackage{xspace}
\usepackage{natbib}
\usepackage{hyperref}
\usepackage{url}

\usepackage{graphicx}
\usepackage{subfigure}
\usepackage{multirow}
\usepackage{multicol}
\usepackage{pifont}
\usepackage{enumitem}
\usepackage{amsmath}
\usepackage{appendix}

\usepackage{tikz}
\newcommand*{\circled}[1]{\lower.7ex\hbox{\tikz\draw (0pt, 0pt)%
    circle (.5em) node {\makebox[1em][c]{\small #1}};}}

\usepackage{amsmath}
\usepackage{amssymb}
\usepackage{mathtools}
\usepackage{amsthm}
\usepackage{listings} %
\newcommand{\zkc}[1]{\textcolor{cyan}{}}
\newcommand{\jjx}[1]{\textcolor{red}{}}
\newcommand{\zj}[1]{\textcolor{pink}{}}

\newcommand{\rev}[1]{{#1}}

\title{CodeDPO: Aligning Code Models with Self Generated and Verified Source Code}

\author{\textbf{Kechi Zhang}$^{1,2}$\footnotemark[2],  \textbf{Ge Li}$^{1,2}$\footnotemark[1],  \textbf{Yihong Dong}$^{1,2}$\footnotemark[2], \textbf{Jingjing Xu}$^{3}$, \\ \textbf{Jun Zhang}$^{3}$,  \textbf{Jing Su}$^{3}$, \textbf{Yongfei Liu}$^{3}$, \\ 
\textbf{Zhi Jin}$^{1,2}$\footnotemark[1] \\
$^1$Key Lab of High Confidence Software Technology (PKU), Ministry of Education \\
$^2$School of Computer Science, Peking University, China \\
$^3$ByteDance \\
\texttt{\{zhangkechi,lige,zhijin\}@pku.edu.cn}}

\begin{document}
\maketitle
\renewcommand{\thefootnote}{\fnsymbol{footnote}}
\footnotetext[2]{Work done during Kechi's internship at ByteDance.}
\footnotetext[1]{Ge Li and Zhi Jin are the corresponding authors.}
\renewcommand{\thefootnote}{\arabic{footnote}}
\begin{abstract}
Code generation models have shown significant potential for programming tasks. 
However, existing training methods like supervised fine-tuning face key limitations: they do not effectively teach models to prioritize correct over incorrect solutions in ambiguous situations, nor do they effectively optimize the runtime efficiency of the generated code. 
To address these challenges, we propose CodeDPO, a framework that integrates preference learning into code generation to improve two key code preference factors: code correctness and efficiency.
CodeDPO employs a novel dataset construction method, utilizing a self-generation-and-validation mechanism that simultaneously generates and evaluates code and test cases. 
The underlying assumption is that test cases executable by multiple code snippets provide more reliable validation, and code that passes more tests is more likely to be correct. 
Through this self-validation process, our PageRank-inspired algorithm iteratively updates the ranking score of each code snippet, ultimately creating a code preference optimization dataset based on correctness and efficiency.
CodeDPO is flexible and scalable, generating diverse preference optimization data without depending on \rev{powerful models such as GPT-4}. 
Through comprehensive evaluations of five widely used benchmarks, CodeDPO demonstrates significant improvements in correctness and efficiency compared to existing methods. 
Our experiments prove that CodeDPO enhances the capabilities of LLMs in code generation and provides a robust foundation for conducting code preference optimization in more complex and challenging real-world scenarios.
\footnote{Code and additional details are available: \url{https://anonymous.4open.science/r/CodeDPO/}}
\end{abstract}


\maketitle

\section{Introduction}


In recent years, code generation models have gained significant attention for their potential to automate software development \cite{zhang2024codeagent}. Models such as GPT-4 \citep{GPT-4}, Claude, and open-source alternatives like Phi \citep{gunasekar2023textbooks,abdin2024phi}, DeepSeekCoder \citep{guo2024deepseek}, and StarCoder \citep{li2023starcoder,lozhkov2024starcoder} have demonstrated the capability of LLMs to handle complex code generation tasks \cite{zhang2024codeagent,zhang2024hirope}. However, one of the ongoing challenges lies in boosting the correctness and efficiency of the generated code. 

\begin{figure}[h]
\centering
  \includegraphics[width=0.8\columnwidth]{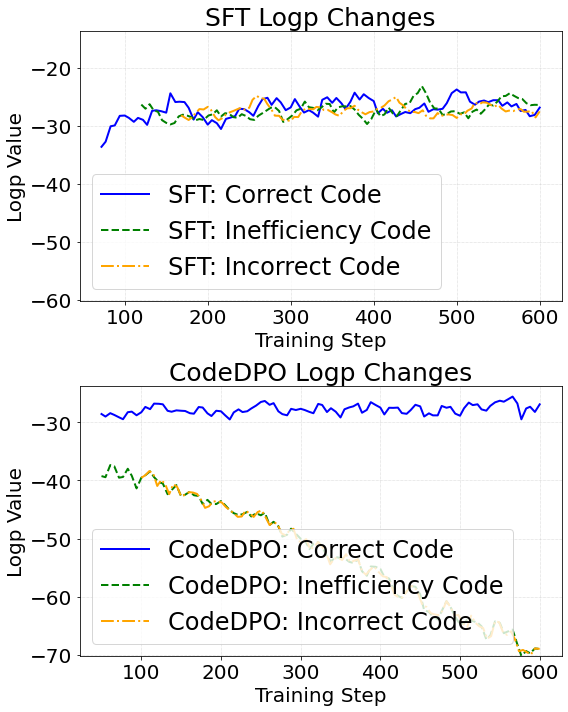}  
\caption{Log probabilities for code with varying correctness and efficiency during Phi-2-2.7B model training on our constructed dataset. The traditional SFT strategy struggles to teach models to prefer correct solutions over incorrect or slow ones. In contrast, our CodeDPO approach effectively optimizes for both correctness and efficiency.}
\label{fig:motivation}
\vspace{-10pt}
\end{figure}

To improve code generation models, a common approach is supervised fine-tuning (SFT) \citep{zhang2023instruction}, where models are trained on pairs of instructions and correct code snippets. 
While SFT improves the overall quality of the generated code, it falls short in teaching models to consistently prefer correct solutions over incorrect ones \citep{hong2024orpo}. 
Figure \ref{fig:motivation} illustrates the likelihood of generating code with varying correctness and efficiency during SFT training. When we adopt SFT training on those correct solutions, as the likelihood of preferred outputs increases, the probability of generating undesirable outputs also rises, leading to performance saturation. 


To address these limitations, recent research has turned to direct preference optimization (DPO) \citep{rafailov2024direct}, a method designed for alignment based on pairwise preference data. 
DPO allows models to rank different outputs and choose preferable solutions (e.g., more factual or helpful). 
While DPO has shown success in reasoning tasks like mathematics \citep{lai2024step,wu2024self}, its application in code generation remains under-explored. Unlike natural language tasks, code generation requires objective metrics, such as executability, which poses challenges for directly applying DPO.
\textbf{In this paper, we first define code preference based on two key factors—Correctness and efficiency.} Correctness refers to whether the code solves the problem accurately, while efficiency measures how quickly the code runs.
Existing methods \citep{codeoptimise,plum} rely heavily on high-quality test cases to assess correctness. However, these approaches struggle to fully address correctness and efficiency, facing limitations such as restricted data diversity, an imbalance between positive and negative samples, and insufficient focus on optimizing code efficiency.
\jjx{Here you can directly define correctness and efficiency as code preference, and then introduce how related work is struggle to achieve this} 

In this paper, we introduce CodeDPO, a novel framework that integrates preference learning into code model training to optimize both correctness and efficiency. 
CodeDPO constructs the dataset from real-world code repositories using a self-generation-and-validation mechanism, where code and test cases are simultaneously generated and evaluated. 
We assume that tests executable by more code snippets are more reliable, and code that passes more tests is more likely to be correct. To implement this, CodeDPO uses a mutual verification process: each receives an initial self-validation score, which is iteratively updated using a PageRank-inspired \citep{page1999pagerank} algorithm. This algorithm adjusts the credibility of each code snippet and tests by considering their relations in cross-verification, prioritizing solutions based on correctness and efficiency. The final preference-optimized dataset is then used to train various code models using the DPO learning algorithm. 
\jjx{You can describe more on the details of the self-verified process.}
A key advantage of CodeDPO is its flexibility. Unlike existing methods that rely on high-quality test cases or powerful models to generate them, CodeDPO does not depend on these resources.
Its self-generation and validation mechanism supports the scalable creation of diverse and robust preference optimization data. This allows our framework to optimize code models for real-world scenarios where high-quality test data may be sparse. 


CodeDPO can serve as a crucial step in the post-training phase of code models. We conduct experiments on five popular benchmarks such as HumanEval \citep{chen2021evaluating}, HumanEval+ \citep{liu2024your}, MBPP \citep{austin2021program}, MBPP+, and DS-1000 \citep{lai2023ds} with CodeDPO, demonstrating its superiority over existing methods.
Notably, we develop a top-performing 6.7B model by building on an existing SFT strategy \citep{guo2024deepseek,wei2023magicoder} and further enhancing it with our CodeDPO approach, achieving an impressive 83.5\% pass rate on HumanEval.
We also conduct ablation studies to investigate the impact of our self-generation-and-validation mechanism and other preference optimization settings. 
Our findings confirm that CodeDPO enhances the code generation capabilities of LLMs while providing a solid foundation for further research into optimizing code generation for both correctness and efficiency.



\section{Related Work}

\subsection{Large Language Models for Code}

Code generation, which automates writing source code from natural language (NL) descriptions, is gaining significant attention. LLMs have shown strong capabilities in this area due to their large-scale training on diverse datasets, such as OpenAI's GPT-4 \citep{GPT-4}, StarCoder \citep{li2023starcoder,lozhkov2024starcoder}, Code Llama \citep{roziere2023code}, and DeepSeekCoder \citep{guo2024deepseek}. 
These models are often fine-tuned further, such as through instruction-supervised fine-tuning (SFT), to maximize their coding potential.
Since gathering high-quality data is difficult, researchers use self-instruct methods to generate synthetic instruction data from powerful models like GPT-4 \citep{wang2022self,alpaca,codealpaca}. Evol-Instruct \citep{luo2023wizardcoder} uses more complex prompts for better data generation. OSS-instruct \citep{wei2023magicoder} allows LLMs to get inspired from real-world code snippets for better quality in coding tasks.
While these SFT methods boost code quality, it does not fully train models to prefer correct solutions over incorrect ones \citep{hong2024orpo}. 
Updating training strategies is critical for improving these code models to handle various coding tasks.

\subsection{Preference Optimization for Code Models}

Preference optimization techniques have recently been used to help LLMs prefer better outputs over weaker ones in various natural language tasks \citep{rafailov2024direct}. 
\rev{The \textbf{Direct Preference Optimization \citep{rafailov2024direct}} has been widely applied to LLM alignment due to its convenience and effectiveness. Its objective is defined as:}
\[
\begin{aligned}
L_{\text{DPO}} = -\mathbb{E}_{(x, y_w, y_l) \sim \mathcal{D}} \left[\log \sigma \left(\beta \log \frac{\pi_{\theta}(y_w \mid x)}{\pi_{\text{ref}}(y_w \mid x)} \right.\right. \\
\left.\left. - \beta \log \frac{\pi_{\theta}(y_l \mid x)}{\pi_{\text{ref}}(y_l \mid x)} \right)\right]
\end{aligned}
\]
\rev{Compared with the SFT loss, the DPO loss introduces a preference-based mechanism. Instead of merely maximizing the likelihood of ground truth data, as in SFT, DPO optimizes the model to align with human preferences by leveraging both preferred responses ($y_w$, winning) and dispreferred responses ($y_l$, losing).}
While DPO has proven effective in reasoning tasks like mathematics \citep{lai2024step}, its use in code generation is still under-explored. 
Code generation requires objective measures of correctness and efficiency, unlike natural language tasks where preferences are often more subjective.
Some works have simply explored PO \cite{codeoptimise,plum,ethayarajh2024kto,zhang2025focused}. 
Code-Optimize \citep{codeoptimise} builds its dataset from the MBPP-train subset, which includes just 384 problems.
PLUM uses GPT-4 to generate tests, which are then used to validate and rank code solutions. PLUM currently achieves state-of-the-art performance in preference optimization for code models.
However, PLUM \citep{plum} faces some limitations. It uses a limited number of tests to validate the code, and the resulting dataset is imbalanced due to its validation method, which means it can only use KTO \citep{ethayarajh2024kto} for training. Additionally, PLUM does not consider the code efficiency.
This paper introduces CodeDPO, which does not rely on \rev{external test cases or powerful models for dataset generation}. Our approach uses a self-generation and validation mechanism to create balanced preference pairs, aiming to optimize both correctness and efficiency.




\section{CodeDPO: Self-Verified Performance Optimization Code Generation Framework}


\begin{figure*}[t]
\centering
  \includegraphics[width=\linewidth]{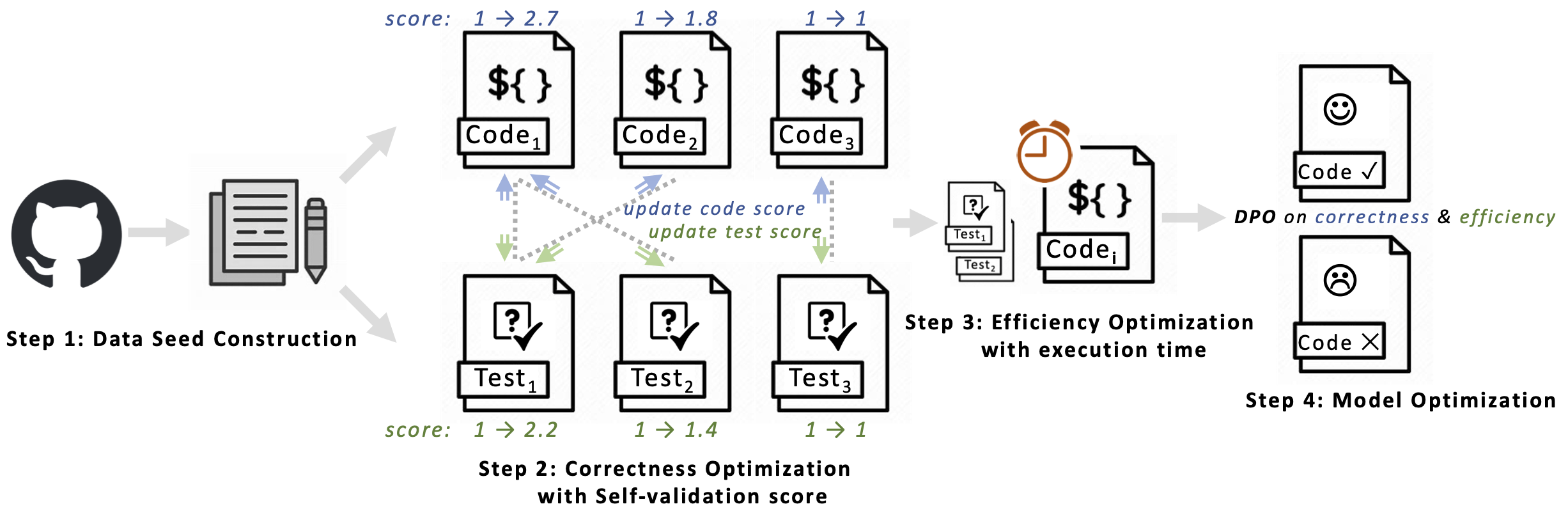}  
\caption{Our CodeDPO involves four steps: \ding{182} Data Seed Construction with real-world source code; \ding{183} Correctness Optimization with self-validation score (in this figure we set $T$ to 2 and $d$ to 0.5. For simplicity, the final score in the figure is rounded to one decimal place. \rev{Details are shown in Appendix \ref{sec:pythonimp}}); \ding{184} Efficiency Optimization with execution time on credible tests; \ding{185} Model Optimization Training. }
\label{fig:method}
\end{figure*}


CodeDPO is designed to integrate preference learning into code generation models, improving both the correctness and efficiency of the generated code. 
As shown in Figure \ref{fig:method}, our method involves four key steps: 
\ding{182} \textbf{Data Seed Construction with real-world source code}: We first collect a data seed from open-source code repositories and generate programming task prompts. 
\ding{183} \textbf{Correctness Optimization with self-validation score}:
We generate code and tests simultaneously, using a self-generation-and-validation loop to build a dataset for correctness optimization. The self-validation score is iteratively updated based on whether the generated code passes the tests. We assume that tests executable by multiple code snippets are more reliable, and code that passes more tests is more likely to be correct. As illustrated in the figure, after two iterations, the score of \textit{code-1} changes from \textit{1} to \textit{1.75} to \textit{2.6875} ($\sim$\textit{2.7} in the figure), as it passes more reliable tests and receives higher scores with each update, indicating a greater likelihood of correctness. 
\ding{184} \textbf{Efficiency Optimization with execution time}: We measure execution time on selected credible test sets to build the dataset for efficiency optimization. In the figure, we select \textit{test-1} and \textit{test-2} as the credible test set to measure the execution time of each code snippet.
\ding{185} \textbf{Model Optimization Training}: We collect the dataset from the previous two stages and use Direct Preference Optimization (DPO) to train various code models.

\subsection{Data Seed Construction}

The data seed construction for CodeDPO is the first step for initiating the preference learning process to generate programming task prompts. We adopt a method inspired by OSS-instruct \citep{wei2023magicoder,wei2024selfcodealign}\footnote{\rev{We follow the implementation provided at \url{https://github.com/bigcode-project/starcoder2-self-align/tree/fd0af77e2773b14696c7cea02a472f9e99d9c4e3}.}}, which extracts key programming concepts from open-source code repositories. These concepts serve as the foundation for generating various programming task prompts.
For example, a code snippet that performs sorting operations might highlight concepts such as sorting algorithms, data structure traversal, and time complexity. From these concepts, we generate code generation prompts. The data seed thus allows the model to explore a wide range of scenarios.

\subsection{Correctness Optimization with Self-Generation and Validation}

Central to CodeDPO is the self-generation-and-validation loop, which enables the model to iteratively update the code correctness rank through mutual validation of code and test cases \citep{chencodet,chen2023teaching,zhang2023self}. 
The process begins by generating multiple candidate code snippets based on a prompt. Simultaneously, corresponding test cases are generated to evaluate these snippets. The validation loop follows these steps:
\textbf{1. Code Generation:} Given an instruction, the model generates a set of candidate code snippets \( C = \{c_1, c_2, ..., c_n\} \).
\textbf{2. Test Case Generation:} Test cases \( T = \{t_1, t_2, ..., t_m\} \) are generated in parallel to validate the candidate snippets.
\textbf{3. Validation Process:} Each code snippet is executed against the generated test cases. The validation outcomes are used to update the self-validation scores for both the code snippets and the test cases.

\paragraph{Ranking Code Snippets and Test Cases Using Self-Validation Scores}

To rank both code snippets and tests, we employ a PageRank-inspired \citep{page1999pagerank} iterative algorithm. Initially, each code and test is assigned a self-validation score of 1. Over a fixed number of iterations \( T = 10 \), these scores are updated based on the performance of the snippets and test cases during validation. 

The self-validation score for code snippets and test cases is updated using the following formulas:

\begin{equation}
\resizebox{\linewidth}{!}{$
\begin{split}
\text{Score}_t(c_i) &= (1 - d) \times \text{Score}_{t-1}(c_i) \\
&\quad + d \times \sum_{t_j} \text{Score}_{t-1}(t_j) \times \text{Link}(t_j, c_i)
\\
\text{Score}_t(t_j) &= (1 - d) \times \text{Score}_{t-1}(t_j) \\
&\quad + d \times \sum_{c_i} \text{Score}_{t-1}(c_i) \times \text{Link}(c_i, t_j)
\end{split}
$}
\end{equation}

Where \( d \) is the damping factor, and \( \text{Link}(t_j, c_i) \) indicates whether a code snippet \( c_i \) passes the test case \( t_j \). This iterative process is repeated until convergence. After \( T \) iterations, the final rankings reflect the quality of the code snippets and test cases based on the correctness.

\subsection{Execution Efficiency Optimization}

In addition to ensuring correctness, CodeDPO integrates execution efficiency optimization to ensure that our approach generates functionally correct and efficient code. During the self-validation loop, the execution time for each code snippet is recorded. However, not all test cases accurately reflect the efficiency of the code. To address this, we use the top-performing code from the correctness optimization phase as a reference, assuming the test cases it passes are credible. The total execution time for each code snippet is then measured based on the subset of these credible tests.
Code snippets that pass these credible test cases with lower execution times are assigned higher efficiency scores. Finally, we collect both fast and slow code snippets as part of the training dataset for execution efficiency optimization, which is used for further training, encouraging the model to prioritize solutions that are accurate and optimized for speed during code generation.

\subsection{Final Dataset and Model Optimization}


The final dataset is built from the previous two optimization dataset construction stages, accounting for correctness and execution time. 
We filter out samples whose ranking scores are identical or too close. 
The final dataset consists of \textbf{93k correctness optimization samples} and \textbf{21k efficiency optimization samples}. 
Each sample includes a unique code problem prompt with a preferred and a rejected solution. 
In the subsequent training, we combine both correctness and efficiency data to optimize the model in both aspects simultaneously.
In our experiments, we apply Direct Preference Optimization (DPO) \citep{rafailov2024direct} across various code models to facilitate optimization learning.
To enhance the stability and robustness of the training process, we employ RPO \citep{pang2024iterative,liu2024provably} format loss, which essentially consists of a weighted SFT loss on the chosen preferences together with the original DPO loss.
We utilize both base models and SFT models as the backbone for further training. 
Our goal is to demonstrate that CodeDPO has the potential to enhance code models at different stages of their training, even for models that have undergone extensive training or fine-tuning.
The setup details are provided in Section \ref{sec:trainingsetting}.

\section{Experiment Setup}
\label{sec:experimentsetup}

In this study, we aim to investigate the following research questions:

\textbf{RQ1: Does CodeDPO improve the correctness of generated code compared to baseline models on standard benchmarks?}
We evaluate on HumanEval \citep{chen2021evaluating}, HumanEval+ \citep{liu2024your}, MBPP \citep{austin2021program}, MBPP+, DS-1000 \citep{lai2023ds} and LiveCodeBench \citep{jain2024livecodebench}. 

\textbf{RQ2: Does CodeDPO enhance the execution efficiency of generated code?}
We measure the execution efficiency of code generated by CodeDPO compared to baseline models.

\textbf{RQ3: What is the impact of the self-generation-and-validation algorithm on CodeDPO's performance?}
We perform ablation studies by removing or modifying the self-generation-and-validation mechanism to assess its contribution to the overall performance.

\textbf{RQ4: How does the choice of preference optimization strategy affect CodeDPO's effectiveness?}
We evaluate different preference optimization strategies, including Direct Preference Optimization (DPO), Kahneman-Tversky Optimization (KTO) \citep{ethayarajh2024kto}, and Supervised Fine-Tuning (SFT), to understand their impact on the model's performance.

\textbf{RQ5: How does data scaling influence the performance of CodeDPO?}
We investigate data scaling by varying the amount of training data to show how data size affects its ability.

\subsection{Backbone LLMs}
\label{sec:setupLLM}

We evaluate several widely used LLMs in the code generation domain for our experiments, covering both \textbf{base models} and \textbf{SFT models} at different training stages. 
For \textbf{base models}, we apply CodeDPO to \textbf{Phi-2 (2.7B)} \citep{gunasekar2023textbooks}, \textbf{DeepSeekCoder-base (1.3B, 6.7B)} \citep{guo2024deepseek}, and \textbf{StarCoder2-base (7B)} \citep{lozhkov2024starcoder}. 
Additionally, we evaluate our method on several fine-tuned \textbf{SFT models} \citep{wei2023magicoder}, including \textbf{Magicoder-CL-7B}, \textbf{Magicoder-S-CL-7B}, \textbf{Magicoder-DS-6.7B}, and \textbf{Magicoder-S-DS-6.7B}, which are fine-tuned based on \textit{CodeLlama-7B} and \textit{DeepSeekCoder-base-6.7B} using state-of-the-art SFT techniques.

While applying the PO phase after SFT is generally recommended \citep{rafailov2024direct}, we extend our evaluation to base models as they can generate more diverse code snippets and offer more significant potential for improvement \citep{wang2024planning}. 
We choose all these popular models as the backbone of our experiments to optimize correctness and execution efficiency.









\subsection{Training and Inference Settings}
\label{sec:trainingsetting}

For dataset construction,  we use \textit{DeepSeekCoder-v2} as the data generation model. 
For each problem prompt, we sample 15 code solutions and test cases from this model with $temperature = 1.5$. 
To construct the preference optimization dataset, we set $T$ to 10 and $d$ to 0.85 for the self-validation score. Our practice shows that this parameter configuration quickly yields a stable ranking score. In this paper, we focus on constructing a Python dataset. \rev{The total cost of our dataset construction process is nearly 80\$.}
For training, we train each code model for 10 epochs and select the best-performing model based on the lowest validation loss. We utilize a learning rate of 5e-6 with a linear scheduler and warm-up. 
For inference, we use greedy search decoding for code generation. We use 16 A100 GPUs for all experiments.

\section{Results and Analyses}


\subsection{Code Correctness (RQ1)}

We evaluate the model performance on five widely-used code generation benchmarks: \textbf{HumanEval}, \textbf{HumanEval+}, \textbf{MBPP}, \textbf{MBPP+}, and \textbf{DS-1000}.
Following the standard training process (base model $\rightarrow$ SFT $\rightarrow$ DPO), we first record the performance of the base model, SFT model, and DPO-aligned model on DeepSeekCoder-6.7B, as shown in Table \ref{tab:dsSFTDPO}. 
With the enhancement of our CodeDPO, the final model achieves an 83.5\% pass rate on HumanEval. Notably, even after high-quality SFT training, CodeDPO still achieves additional performance improvements. CodeDPO plays a crucial role in the post-training phase of code models, significantly boosting overall performance.

\begin{table*}[h!]
\centering
\begin{tabular}{l|c|c|c|c}
\toprule
\textbf{Model} & \textbf{HumanEval} & \textbf{HumanEval+} & \textbf{MBPP} & \textbf{MBPP+} \\
\midrule
DeepSeekCoder-6.7B-base & 47.60 & 39.60 & 70.20 & 56.60 \\
+ SFT \textit{(with MagiCoder-OSS-instruct)} & 73.17  & 68.29  & 76.72 & 66.67 \\
+ SFT + \textbf{Our CodeDPO} & \textbf{83.54}  & \textbf{76.22}  & \textbf{80.70} & \textbf{70.93} \\
\bottomrule
\end{tabular}
\caption{Pass rates (\%) of code models at different stages on HumanEval(+) and MBPP(+). We track the performance of the base model, SFT model, and DPO-aligned model on DeepSeekCoder-6.7B. Our CodeDPO shows additional improvements, even after high-quality SFT training.}
\label{tab:dsSFTDPO}
\end{table*}


We further evaluate the performance of CodeDPO alongside baselines\footnote{The baselines have not yet published their datasets. We reproduced the Code-Optimize experiment based on the reported settings. For PLUM, we report results from their paper using models identical to ours, which is why some models do not include PLUM results.} on a wide range of base models and SFT models.
As shown in Table \ref{tab:hembpp}, CodeDPO achieves the best performance on both HumanEval(+) and MBPP(+). Compared to the baseline models in the first row of each block, we observe that CodeDPO delivers significant improvements across all models, regardless of their initial performance. Notably, we achieve a 36.1\% relative improvement on StarCoder2-7B. Additionally, CodeDPO shows remarkable gains on the more challenging HumanEval+, demonstrating its robustness under stricter evaluation.
Thanks to CodeDPO's data construction strategy, we can build a reliable preference dataset that helps the model favour high-quality outputs, leading to more robust and reliable code generation.



For DS-1000, as shown in Table \ref{tab:ds1000} and \ref{tab:ds1000full}, we further evaluate CodeDPO across different libraries. We did not incorporate prior knowledge of specific Python libraries in our data construction. While we observe slight performance drops in the Torch and TensorFlow settings, this may be due to the relatively low percentage of these libraries in our dataset construction. However, CodeDPO demonstrates overall performance improvements over their respective baselines.
DS-1000 differs from benchmarks like HumanEval and MBPP in data format and the coding skills it assesses, and it ensures that \textit{it is excluded from nearly all models' training sets}. It proves that CodeDPO can enhance the model's coding capabilities in more complex and diverse scenarios. Details are in Appendix \ref{sec:rqds1000}.

We conducted additional experiments on \textbf{LiveCodeBench} \citep{jain2024livecodebench} in Table \ref{tab:livecodebench_results}, one of the most challenging benchmarks for real-world competitive coding tasks. The results indicate that CodeDPO demonstrates significant performance improvements for both the base model and the supervised fine-tuning (SFT) model across all difficulty levels. The gains are particularly notable in the "medium" and "hard" subsets, which represent some of the most challenging problems in competitive programming tasks. These findings highlight the effectiveness of the proposed framework for real-world, complex coding tasks.

\begin{table}[h!]
\centering
\resizebox{\linewidth}{!}{
\begin{tabular}{l|c|c}
\toprule
\textbf{Model} & \textbf{HumanEval (/+)} & \textbf{MBPP (/+)} \\
\midrule
\textbf{\textit{SFT Model}}  \\
\midrule
MagiCoder-CL-7B & 51.21 / 48.78 & 65.60 / 55.82 \\
Our CodeDPO & \textbf{60.36 / 54.87} & \textbf{70.93 / 59.15} \\
\textit{Code-Optimise}  & 48.78 / 46.95 & 67.17 / 57.14 \\
\midrule
MagiCoder-S-CL-7B & 67.07 / 61.59 & 69.58 / 60.58 \\
Our CodeDPO & \textbf{74.39 / 71.95} & \textbf{71.43 / 61.40} \\
\textit{Code-Optimise}  & 64.63 / 54.88 & 69.42 / 60.15 \\
\textit{PLUM}           & 73.80 / 69.50 & 71.40 / 60.80 \\
\midrule
MagiCoder-DS-6.7B & 57.93 / 53.66 & 75.93 / 64.02 \\
Our CodeDPO & 67.07 / 62.80 & \textbf{81.70 / 68.92} \\
\textit{Code-Optimise}  & 57.93 / 51.83 & 76.19 / 64.91 \\
\textit{PLUM}           & \textbf{71.30 / 65.90} & 79.60 / 66.70 \\
\midrule
MagiCoder-S-DS-6.7B & 73.17 / 68.29 & 76.72 / 66.67 \\
Our CodeDPO & \textbf{83.54 / 76.22} & \textbf{80.70 / 70.93} \\
\textit{Code-Optimise}  & 68.90 / 64.63 & 78.20 / 67.92 \\
\textit{PLUM}           & 80.50 / 73.80 & 80.40 / 69.30 \\
\midrule
\textbf{\textit{Base Model}}  \\
\midrule
Phi-2-2.7B & 48.78 / 46.34 & 65.34 / 54.49 \\
Our CodeDPO & \textbf{57.32 / 51.83} & \textbf{69.05 / 56.88} \\
\textit{Code-Optimise}  & 49.39 / 47.56 & 67.42 / 55.80 \\
\midrule
DeepSeekCoder-1.3B & 31.53 / 28.65 & 57.40 / 48.67 \\
Our CodeDPO & \textbf{42.07 / 38.04} & \textbf{61.37 / 53.43} \\
\textit{Code-Optimise}  & 34.15 / 30.49 & 59.15 / 49.87 \\
\midrule
DeepSeekCoder-6.7B & 47.60 / 39.60 & 70.20 / 56.60 \\
Our CodeDPO & \textbf{59.75 / 51.83} & 72.18 / \textbf{60.01} \\
\textit{Code-Optimise}  & 47.56 / 37.20 & 72.18 / 57.64 \\
\textit{PLUM}           & 56.70 / 48.80 & \textbf{72.90} / 58.90 \\
\midrule
StarCoder2-7B & 35.40 / 29.90 & 54.40 / 45.60 \\
Our CodeDPO & \textbf{48.17 / 34.15} & 58.40 / \textbf{49.37} \\
\textit{Code-Optimise}  & 32.32 / 28.05 & 58.90 / 47.89 \\
\textit{PLUM}           & 46.30 / \textbf{39.60} & \textbf{60.40} / 49.10 \\
\bottomrule
\end{tabular}
}
\caption{Pass rate (\%) of CodeDPO compared to baseline models on HumanEval and MBPP.}
\label{tab:hembpp}
\end{table}


\begin{table}[h!]
\centering
\resizebox{\linewidth}{!}{
\begin{tabular}{l|c|c|c|c|c|c|c|c}
\toprule
\textbf{Model} & {{plot}} & {{np}} & {{pd}} & {{torch}} & {{scipy}} & {{sk }} & {{tf }} & \textbf{Avg} \\
\midrule
Magic-CL-7B & 54.8 & 16.4 & 16.5 & 17.6 & 23.6 & 29.6 & \textbf{33.3} & 25.5 \\
Our CodeDPO      & \textbf{57.4} & \textbf{37.3} & \textbf{22.7} & \textbf{22.1} & \textbf{35.8} & \textbf{31.3} & 31.1 & \textbf{34.0} \\
\midrule
Magic-DS-6.7B & 55.5 & 37.7 & 28.2 & \textbf{25.0} & 34.0 & \textbf{45.2} & \textbf{33.3} & 37.1 \\
Our CodeDPO      & \textbf{59.4} & \textbf{40.5} & \textbf{29.2} & 23.5 & \textbf{39.6} & 42.6 & 31.1 & \textbf{38.7} \\
\midrule
\midrule
Phi-2-2.7B       & 42.6 & 33.6 & 15.5 & \textbf{16.2} & 17.0  & 11.3 & \textbf{17.8} & 23.5 \\
Our CodeDPO      & \textbf{49.0}  & \textbf{33.6} & \textbf{16.5} & 14.7 & \textbf{20.8} & \textbf{14.8} & 13.3 & \textbf{25.3} \\
\midrule
StarCoder2-7B & 54.2 & 37.7 & 18.6 & \textbf{25.0}  & 31.1 & 23.5 & \textbf{35.6} & 31.4 \\
Our CodeDPO      & \textbf{56.8} & \textbf{38.2} & \textbf{18.9} & 20.6 & \textbf{39.6} & \textbf{25.2} & 31.1 & \textbf{32.6} \\
\bottomrule
\end{tabular}
}
\caption{Pass rate (\%) of CodeDPO on DS-1000 across seven libraries using greedy decoding.}
\label{tab:ds1000}
\end{table}



\begin{table}[h!]
    \centering
    \resizebox{0.8\linewidth}{!}{
    \begin{tabular}{lccc}
        \toprule
        \textbf{Model} & \textbf{Easy} & \textbf{Medium} & \textbf{Hard} \\
        \midrule
        \textbf{Base Model} & & & \\
        DeepSeek-Coder-6.7B & 39.9 & 7.4 & 0.4 \\
        Our CodeDPO & 51.9 & 12.2 & 0.7 \\
        \midrule
        \textbf{SFT Model} & & & \\
        MagiCoder-S-DS-6.7B & 48.1 & 10.7 & 0.1 \\
        Our CodeDPO & 53.1 & 16.3 & 0.7 \\
        \bottomrule
    \end{tabular}
    }
    \caption{\rev{Performance comparison on LiveCodeBench across difficulty levels.}}
    \label{tab:livecodebench_results}
\end{table}

\subsection{Code Efficiency (RQ2)}

\begin{figure}[h]
\centering
  \subfigure[HumanEval+]{
    \begin{minipage}[b]{0.8\linewidth}
      \centering
      \includegraphics[width=\linewidth]{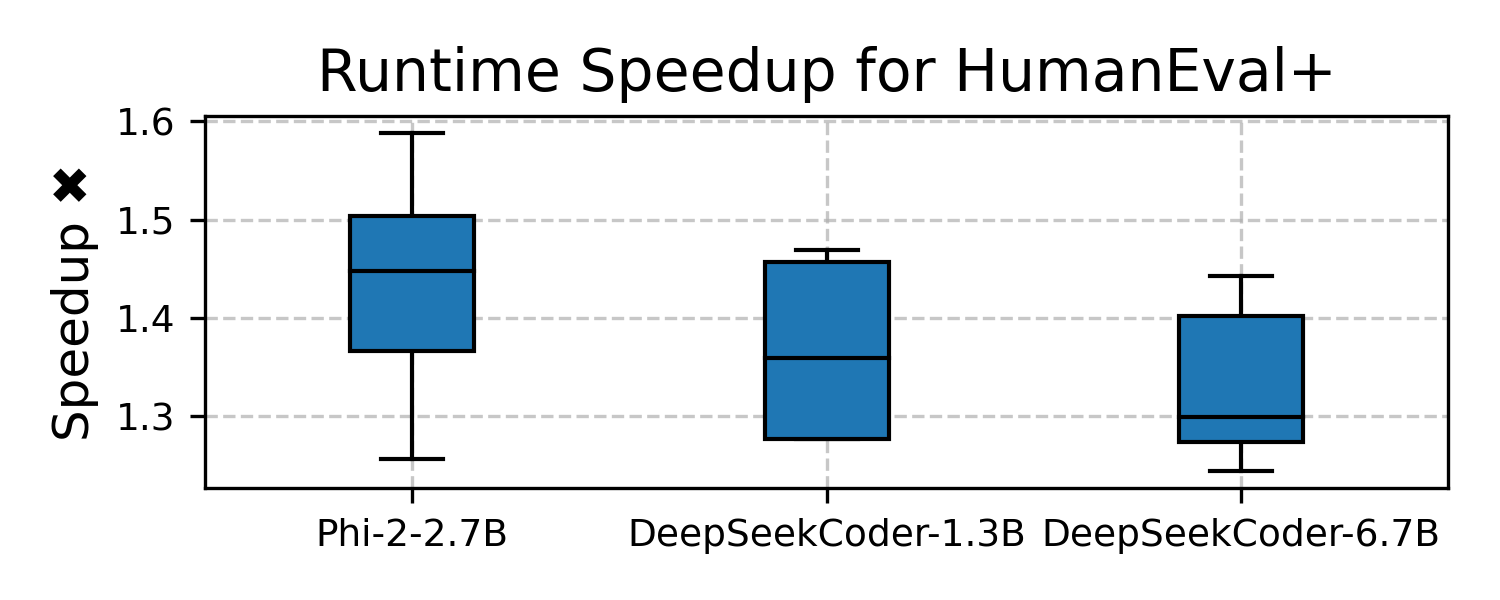} \\
      \includegraphics[width=\linewidth]{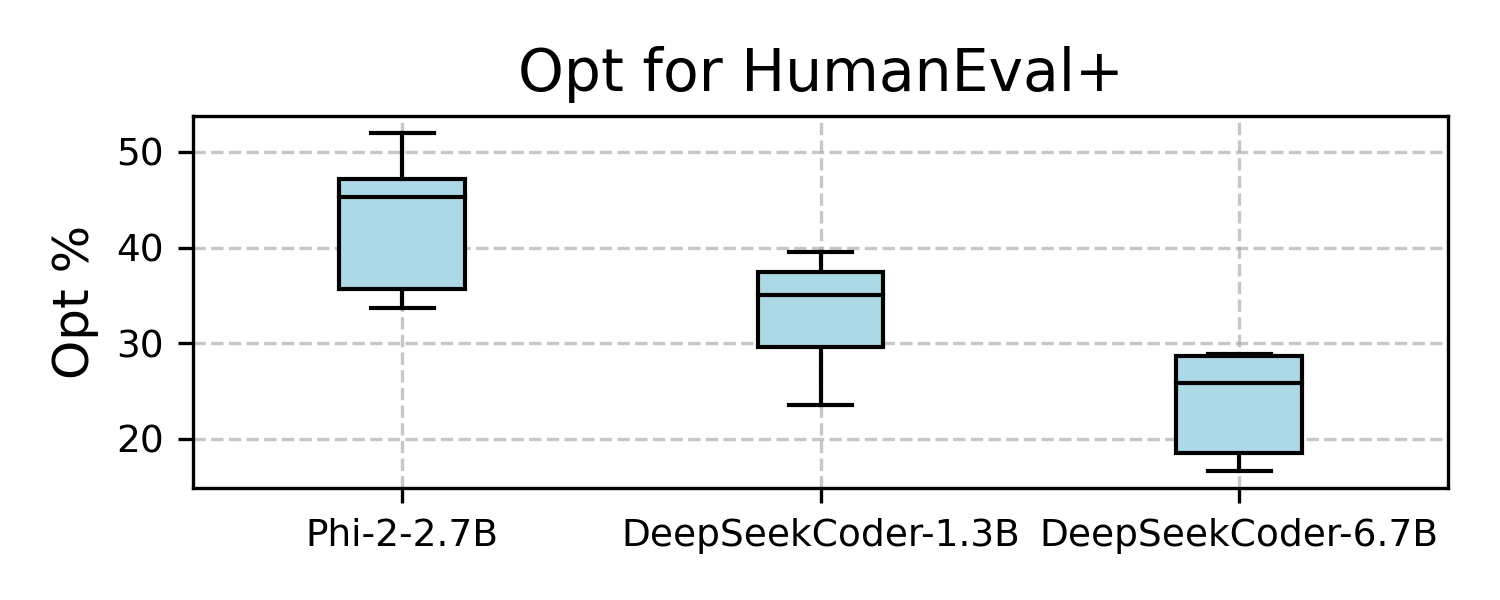}
    \end{minipage}
  }
  \subfigure[MBPP+]{
    \begin{minipage}[b]{0.8\linewidth}
      \centering
      \includegraphics[width=\linewidth]{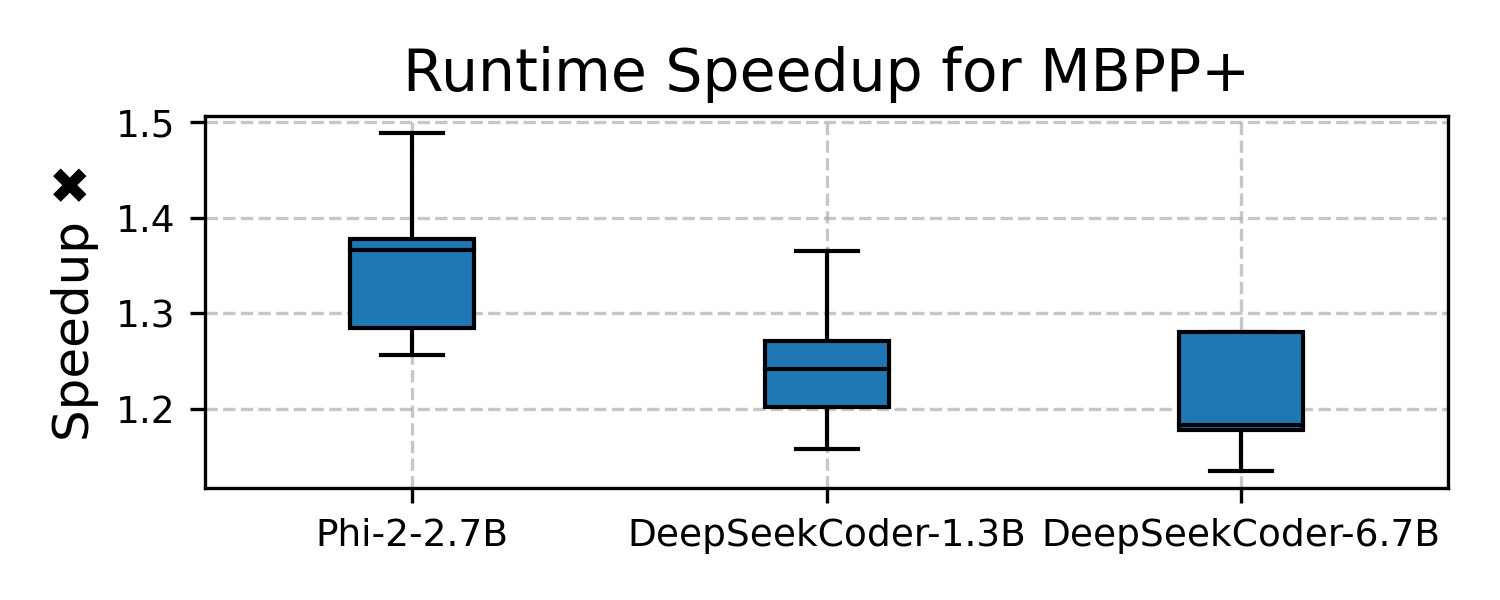} \\
      \includegraphics[width=\linewidth]{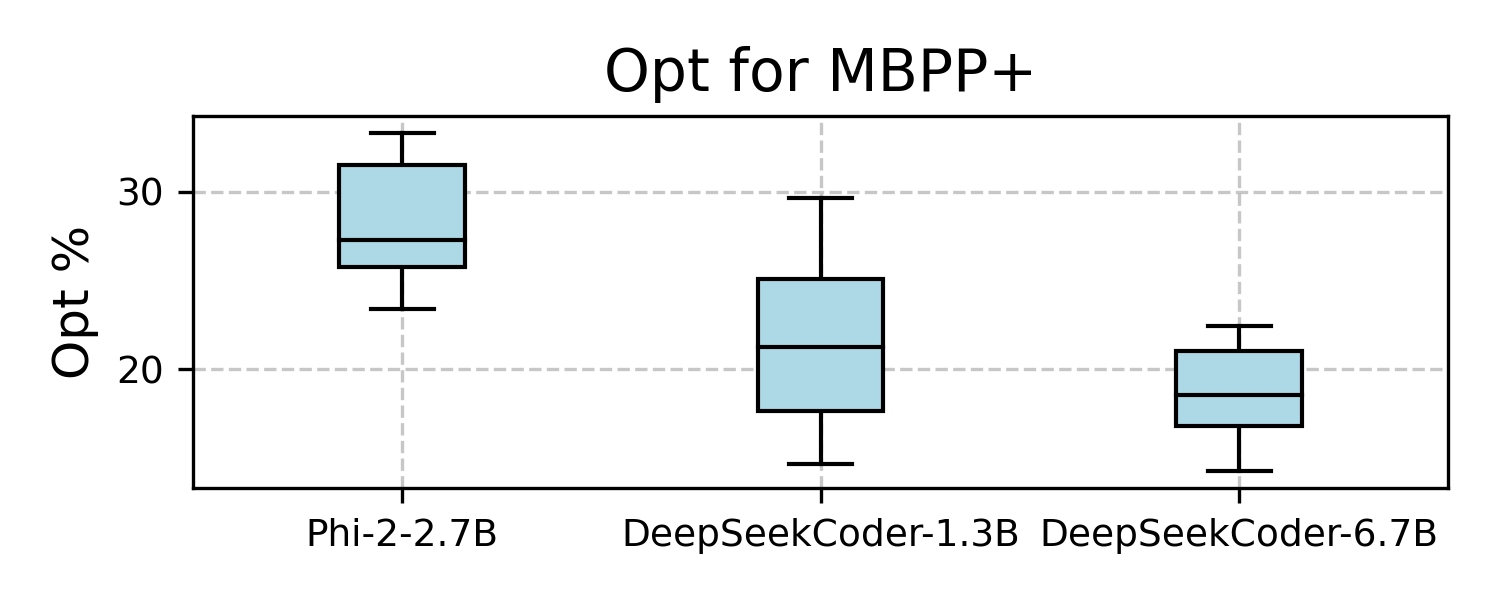}
    \end{minipage}
  }
\caption{Runtime Speedup and Percentage of Optimized Code on HumanEval+ and MBPP+.}
\label{fig:speedupcomparison}
\end{figure}

To address \textbf{RQ2}, we follow existing methods \citep{shypulalearning} by measuring the execution time of the generated code and calculating the \textbf{speed-up} ratio. We also evaluate the \textbf{percentage of optimized code} before and after applying CodeDPO, where a program is considered optimized if it is at least 10\% faster than its baseline. These metrics are based on the intersection of solved problems before and after applying CodeDPO. We select \textbf{HumanEval+} and \textbf{MBPP+} for evaluation because they significantly expand the diversity of test cases, making them more reliable for measuring the execution efficiency of the generated code under a variety of edge cases.
Since runtime environments can affect measurements, we repeat each evaluation five times and show the distribution in Figure \ref{fig:speedupcomparison}. It is clear that CodeDPO consistently improves code performance. The speed-up ratio shows that our method speeds up the code by 1.25 to 1.45 times. 
Additionally, the percentage of optimized code indicates that after applying CodeDPO, around 20\%-45\% of generated code solutions have been improved, confirming its effectiveness in enhancing code efficiency.



\subsection{Ablation Studies}

\subsubsection{Self Generation and Validation Algorithm (RQ3)}
\label{sec:experimentAlg}


\paragraph{Correlation between self-validation scores and actual code accuracy using HumanEval ground truth tests} 
To evaluate the effectiveness of our self-generation-and-validation algorithm, we examine the correlation between self-validation scores and actual code accuracy. We use HumanEval for \textit{this preliminary experiment}. 
For each problem in HumanEval, we sample 15 code solutions and tests following the setting in Section \ref{sec:experimentsetup}, and then use different strategies to rank these generated codes. To evaluate the rank quality, we execute with the ground truth for each code to get the actual code accuracy. 
We consider three experimental strategies: \ding{182} \textbf{Self-validation score}, which refers to our original method. \ding{183} \textbf{Filter with all tests}, which assumes all generated test cases are correct and uses them to judge code correctness. This approach creates passed/non-passed pairs, similar to the baseline \textbf{PLUM} (though PLUM uses GPT-4 for test generation, while we use a more cost-effective model). \ding{184} \textbf{Sort by number of passed tests}, which counts the number of passed tests for each code among all generated tests, using the code with the most and least passed tests as the comparison pair. \rev{It is commonly employed in post-processing methods, such as \textbf{CodeT} \cite{chencodet}.}

Table \ref{tab:correlation} presents the Spearman, Kendall's Tau, and Normalized Discounted Cumulative Gain (NDCG) metrics for the different ranking strategies. Our experiments show that the self-validation score is highly correlated with actual code accuracy.
In contrast, filtering by all tests heavily depends on the quality of the test generation model. Sorting by the number of passed tests treats all tests equally important. However, due to the inherent uncertainty in generated tests, these methods can be vulnerable to low-quality tests. Our proposed self-validation method employs a mutual reinforcement mechanism to update the credibility of both code and tests, effectively mitigating these issues.

\begin{table}[h!]
\centering
\resizebox{\linewidth}{!}{
\begin{tabular}{l|c|c|c}
\toprule
\textbf{Method} & \textbf{Spearman} & \textbf{Kendall's Tau} & \textbf{NDCG} \\
\midrule
Self-validation score & \textbf{0.8598} & \textbf{0.8047} & \textbf{0.9653} \\
\midrule
Filter with all tests & 0.6114 & 0.6114 & 0.8753 \\
Sort by \# of passed tests & 0.7724 & 0.7250 & 0.9162 \\
\bottomrule
\end{tabular}
}
\caption{Correlation between self-validation score and actual code accuracy on HumanEval.}
\label{tab:correlation}
\end{table}

\paragraph{Impact of self-validation score on model performance} We apply these strategies to construct datasets and evaluate the final model performance in code generation. Table \ref{tab:ablationonscorecombine} presents the model performance across various dataset construction strategies. We introduce a new strategy—\textit{random selection}—which randomly selects two code solutions from the generated code as the preference pair. The experiment results demonstrate that the self-generation-and-validation algorithm plays an essential role in ensuring the correctness and reliability of the preference dataset construction. Details are shown in Appendix \ref{sec:rqstrategies}.

\begin{table}[h!]
\centering
\resizebox{\linewidth}{!}{
\begin{tabular}{l|c|c}
\toprule
\textbf{Model} & \textbf{HumanEval (/+)} & \textbf{MBPP (/+)} \\
\midrule
DeepSeekCoder-1.3B & 31.53 / 28.65 & 57.40 / 48.60 \\
\midrule
\textbf{\textit{Data Construction Strategies}} & & \\
Filter with all tests & 34.75 / 29.89 & 57.40 / 48.80 \\
Sort by \# of passed tests & 37.19 / 31.09 & 58.39 / 50.37 \\
Random selection & 21.34 / 18.29 & 48.94 / 38.35 \\
Our CodeDPO & \textbf{42.07 / 38.04} & \textbf{61.37 / 53.43} \\
\midrule
\textbf{\textit{Training Strategies}} & & \\
SFT & 39.02 / 35.36 & 59.45 / 50.26 \\
Our CodeKTO & 40.85 / 35.98 & 59.65 / 50.13 \\
Our CodeDPO & \textbf{42.07 / 38.04} & \textbf{61.37 / 53.43} \\
\bottomrule
\end{tabular}
}
\caption{Ablation study of performance  based on \textit{DeepSeekCoder-1.3B}.}
\label{tab:ablationonscorecombine}
\end{table}

\subsubsection{Impact of PO Training Strategy (RQ4)}

We explore the impact of different preference optimization strategies (DPO, KTO, and SFT) on model performance. For training, the SFT strategy uses the best code solution from our constructed dataset. In our KTO strategy, we replace DPO with KTO in our framework.
As shown in Figure \ref{fig:motivation}, the traditional SFT strategy struggles to guide the model in preferring correct solutions over incorrect or slower ones during training. The results in Table \ref{tab:ablationonscorecombine} demonstrate that DPO performs best among these strategies. Benefiting from our dataset construction method, we can obtain well-balanced preference pairs, enhancing the contrastive mechanism in DPO. Details are shown in Appendix \ref{sec:rqkto}.



\subsection{Data Scaling Law for CodeDPO (RQ5)}
To address \textbf{RQ5}, we explore how scaling the training data affects CodeDPO’s performance. 
We show the experiment results for RQ5, which can help us explore how scaling the training data affects CodeDPO’s performance. We train the model with varying amounts of data—25\%, 50\%, and 75\%—and evaluate its impact on the model performance. 
As shown in Table \ref{tab:scaling}, increasing the data consistently improves model performance, but these improvements gradually plateau as the dataset size grows. 
For example, HumanEval scores rise from 32.92 (25\%) to 41.46 (75\%), with similar trends observed on MBPP. 
In our experiments, we carefully balance performance gains and training costs, ensuring optimal results with CodeDPO. 
In further research, we plan to expand the current training scale to explore the extreme limits of CodeDPO’s performance.


\begin{table}[h!]
\centering
\resizebox{\linewidth}{!}{
\begin{tabular}{l|c|c|c|c}
\toprule
\textbf{Model} & \textbf{HumanEval} & \textbf{HumanEval+} & \textbf{MBPP} & \textbf{MBPP+} \\
\midrule
DeepSeekCoder-1.3B-base & 31.53 & 28.65 & 57.40 & 48.67 \\
\midrule
25\% & 32.92 & 29.87 & 55.13 & 47.87 \\
50\% & 36.59 & 31.70 & 58.14 & 49.87 \\
75\% & 41.46 & 37.80 & 60.65 & 52.63 \\
\midrule
Our CodeDPO & 42.07 & 38.04 & 61.37 & 53.43 \\
\bottomrule
\end{tabular}
}
\caption{Model Performances with different data scaling in our CodeDPO.}
\label{tab:scaling}
\end{table}

\subsection{\rev{Overlap Avoidance with Existing Benchmarks}}

\rev{The seed dataset for CodeDPO was randomly selected from the open-source pretraining dataset \emph{The Stack}, consisting of approximately 100k functions.  This design explicitly considers data decontamination, since the seed dataset has already gone through rigorous data decontamination. It suggests that our dataset is unlikely to introduce additional data leakage beyond the seeds. To ensure quality, we applied a simple filtering process using tools like Tree-sitter and Pyright for static analysis and code formatting.}

\rev{We intentionally avoided introducing any prior knowledge that might lead to significant overlap with evaluation benchmarks. 
We also implemented post-sampling data decontamination, similar to MagiCoder and StarCoder. However, given the already low overlap, this process only removed fewer than 30 samples. Thus, we can ensure that there is no risk of the dataset containing examples highly similar to the test sets.
}

\rev{
To assess potential overlap for the final dataset with exisiting benchmarks, we followed the methodology used in MagiCoder. Specifically, we calculated the cosine similarity between HumanEval and the synthetic data generated by different methods. Below are the average similarity scores in Table \ref{tab:similarity_scores}.}

\begin{table}[h!]
    \centering
    \rev{
    \begin{tabular}{lc}
        \toprule
        \textbf{Dataset} & \textbf{Avg Similarity Score} \\
        \midrule
        Self-Instruct & 0.169 \\
        Evol-Instruct & 0.131 \\
        OSS-Instruct & 0.105 \\
        CodeDPO & 0.109 \\
        \bottomrule
    \end{tabular}
    }
    \caption{\rev{Average similarity scores between datasets and HumanEval.}}
    \label{tab:similarity_scores}
\end{table}

\rev{These results demonstrate that CodeDPO has a comparable or even lower overlap with HumanEval than most other widely used datasets, ensuring the validity and reliability of our evaluation.}


\section{Conclusion}

We propose CodeDPO, a preference optimization framework for code models that focuses on both correctness and efficiency. CodeDPO iteratively updates the self-validation score, prioritizing solutions based on correctness and efficiency.
Our work technically validates the reliability of self-validation to synthesize preference optimization data, eliminating the need for complex resources such as pre-existing tests or powerful generation models. We hope this work opens new avenues for synthesizing data and implementing large-scale preference optimization for code models.

\section{Acknowledgments}

This research is supported by the National Key R\&D Program under Grant No. 2023YFB4503801, the National Natural Science Foundation of China under Grant No. 62192733, 62192730, 62192731, 62072007 and the Major Program (JD) of Hubei Province (No. 2023BAA024).

\section*{Limitation}
There are several limitations to our work that we aim to address:


\paragraph{Comparison with Advanced RL Techniques such as DeepSeek-R1}

Although our study showcases the effectiveness of CodeDPO, it does not thoroughly compare this method with other advanced reinforcement learning (RL) alignment techniques like DeepSeek-R1 \cite{guo2025deepseekr1}. Techniques such as GRPO in DeepSeek-R1 are typically designed for online RL alignment, requiring substantial training resources, high-quality datasets, and complex reward environments, which can be prohibitively resource-intensive.
In contrast, CodeDPO focuses on offline alignment methods, using approximations and necessary simplifications to achieve optimization objectives. This allows CodeDPO to deliver results comparable to or even matching those of advanced online RL methods, but with notably reduced resource demands. 
Our self-validation scores also indicate that our ranking methods are robust even with lower-quality source code and tests, demonstrating consistent performance even with less advanced generation models.
Given its low resource requirements and consistent performance, CodeDPO is well-suited to a broad range of code generation scenarios. However, a comprehensive evaluation of how CodeDPO stacks up against these sophisticated RL techniques in terms of performance and efficiency is a potential avenue for future research.

\paragraph{Limitations of Current Correctness Evaluation}
Firstly, constrained by current mainstream correctness evaluation methods, the test-case-driven functional correctness DPO is still not enough for code model. Current methods for evaluating correctness heavily rely on high-quality test cases or powerful models (e.g., GPT-4) to generate reliable outputs. To address these limitations, our paper introduces a self-validation data generation method that reduces dependency on such resources while maintaining robustness.

Because our method does not require high-quality test cases or strong external models, it is well-suited for scaling to larger datasets and can be applied to a wide range of code models. This scalability provides a foundation for improving correctness and efficiency across diverse code tasks.

\paragraph{Incorporating Readability and Security}

Next, beyond correctness and efficiency, incorporating readability and security metrics into our extended CodeDPO framework is a natural extension: Metrics such as comment-to-code ratio, consistent variable naming, and adherence to coding style guides could be integrated into the preference learning process. For instance, LLMs could act as judges to evaluate readability alongside correctness. Techniques like static code analysis and detection of code smell and common vulnerabilities could help identify and penalize insecure patterns during data construction, contributing to safer code generation. We plan to explore these deeper alignment objectives in future work.


\bibliography{main}

\clearpage
\newpage
\appendix

\section{Code Correctness on DS-1000 (RQ1)}
\label{sec:rqds1000}
For DS-1000, as shown in Table \ref{tab:ds1000full}, we further evaluate CodeDPO across different libraries on various backbone models.
\begin{table*}[h!]
\centering
\resizebox{0.8\linewidth}{!}{
\begin{tabular}{l|c|c|c|c|c|c|c|c}
\toprule
\textbf{Model} & {\textbf{\scriptsize plot (155)}} & {\textbf{\scriptsize np (220)}} & {\textbf{\scriptsize pd (291)}} & {\scriptsize \textbf{torch (68)}} & {\scriptsize \textbf{scipy (106)}} & {\scriptsize \textbf{sk (115)}} & {\scriptsize \textbf{tf (45)}} & \textbf{Average} \\
\midrule
\textbf{\textit{SFT Model}}     \\
\midrule
Magic-CL-7B & 54.8 & 16.4 & 16.5 & 17.6 & 23.6 & 29.6 & \textbf{33.3} & 25.5 \\
Our CodeDPO      & \textbf{57.4} & \textbf{37.3} & \textbf{22.7} & \textbf{22.1} & \textbf{35.8} & \textbf{31.3} & 31.1 & \textbf{34.0} \\
\midrule
Magic-S-CL-7B & 52.3 & 43.2 & 30.6 & \textbf{47.1} & 34.9 & \textbf{46.1} & \textbf{44.4} & 40.7 \\
Our CodeDPO      & \textbf{58.7} & \textbf{44.5} & \textbf{31.3} & 38.2 & \textbf{40.6} & 42.6 & 33.3 & \textbf{41.3} \\
\midrule
Magic-DS-6.7B & 55.5 & 37.7 & 28.2 & \textbf{25.0} & 34.0 & \textbf{45.2} & \textbf{33.3} & 37.1 \\
Our CodeDPO      & \textbf{59.4} & \textbf{40.5} & \textbf{29.2} & 23.5 & \textbf{39.6} & 42.6 & 31.1 & \textbf{38.7} \\
\midrule
Magic-S-DS-6.7B & 53.5 & 49.5 & 30.6 & \textbf{47.1} & 35.8 & \textbf{53.0} & \textbf{40.0} & 42.9 \\
Our CodeDPO      & \textbf{59.4} & \textbf{50.5} & \textbf{31.9} & 39.7 & \textbf{41.5} & 47.8 & 33.3 & \textbf{43.7} \\
\midrule
\textbf{\textit{Base Model}} \\
\midrule
Phi-2-2.7B       & 42.6 & 33.6 & 15.5 & \textbf{16.2} & 17.0  & 11.3 & \textbf{17.8} & 23.5 \\
Our CodeDPO      & \textbf{49.0}  & \textbf{33.6} & \textbf{16.5} & 14.7 & \textbf{20.8} & \textbf{14.8} & 13.3 & \textbf{25.3} \\
\midrule
DSC-1.3B & \textbf{36.8} & 19.5 & 10.0   & 14.7 & 10.4 & \textbf{17.4} & \textbf{11.1} & 17.5 \\
Our CodeDPO      & 34.8 & \textbf{23.6} & \textbf{10.7} & \textbf{14.7} & \textbf{20.8} & 13.9 & 8.9 & \textbf{18.9} \\
\midrule
DSC-6.7B & 52.3 & 35.5 & 20.6 & \textbf{19.1} & 24.5 & \textbf{37.4} & \textbf{22.2} & 31.1 \\
Our CodeDPO      & \textbf{56.8} & \textbf{36.4} & \textbf{21.6} & 17.6 & \textbf{34.0}  & 34.8 & 20.0   & \textbf{32.8} \\
\midrule
StarCoder2-7B & 54.2 & 37.7 & 18.6 & \textbf{25.0}  & 31.1 & 23.5 & \textbf{35.6} & 31.4 \\
Our CodeDPO      & \textbf{56.8} & \textbf{38.2} & \textbf{18.9} & 20.6 & \textbf{39.6} & \textbf{25.2} & 31.1 & \textbf{32.6} \\
\bottomrule
\end{tabular}
}
\caption{More results on Pass rate (\%) of CodeDPO on DS-1000 across seven libraries using greedy decoding.}
\label{tab:ds1000full}
\end{table*}

\section{Self Generation and Validation Algorithm (RQ3)}
\label{sec:rqstrategies}
We apply different strategies to construct datasets and evaluate the final model performance in code generation. We conduct experiments on various models and the detailed results are shown in Table \ref{tab:ablationonscoreAPP}.

\begin{table*}[h!]
\centering
\resizebox{0.6\linewidth}{!}{
\begin{tabular}{l|c|c|c|c}
\toprule
\textbf{Model} & \textbf{HumanEval} & \textbf{HumanEval+} & \textbf{MBPP} & \textbf{MBPP+} \\
\midrule
Phi-2-2.7B & 48.78 & 46.34 & 65.34 & 54.49 \\
Our CodeDPO & \textbf{57.32} & \textbf{51.83} & \textbf{69.05} & \textbf{56.88} \\
Filter with all tests & 49.39 & 48.17 & 69.17 & 55.13 \\
Sort by \# of passed tests & 50.60 & 49.39 & 67.16 & 54.88 \\
Random selection & 22.56 & 18.90 & 45.11 & 36.59 \\
\midrule
DeepSeekCoder-1.3B & 31.53 & 28.65 & 57.40 & 48.60 \\
Our CodeDPO & \textbf{42.07} & \textbf{38.04} & \textbf{61.37} & \textbf{53.43} \\
Filter with all tests & 34.75 & 29.89 & 57.40 & 48.80 \\
Sort by \# of passed tests & 37.19 & 31.09 & 58.39 & 50.37 \\
Random selection & 21.34 & 18.29 & 48.94 & 38.35 \\
\bottomrule
\end{tabular}
}
\caption{More results on ablations of our self validation score on the trained model performance.}
\label{tab:ablationonscoreAPP}
\end{table*}

\section{Impact of PO Training Strategy (RQ4)}
\label{sec:rqkto}
We explore the impact of different preference optimization strategies (DPO, KTO, and SFT) on model performance. For training, the SFT strategy uses the best code solution from our constructed dataset. In our KTO strategy, we replace DPO with KTO in our framework. We conduct experiments on various models and the detailed results are shown in Table \ref{tab:ablationdpoAPP}.

\begin{table*}[h!]
\centering
\begin{tabular}{l|c|c|c|c}
\toprule
\textbf{Model} & \textbf{HumanEval} & \textbf{HumanEval+} & \textbf{MBPP} & \textbf{MBPP+} \\
\midrule
Phi-2-2.7B & 48.78 & 46.34 & 65.34 & 54.49 \\
SFT & 55.49 & 49.22 & 66.87 & 55.76 \\
Our CodeDPO & \textbf{57.32} & \textbf{51.83} & \textbf{69.05} & \textbf{56.88} \\
Our CodeKTO & 54.88 & 51.22 & 64.91 & 53.63 \\
\midrule
DeepSeekCoder-1.3B-base & 31.53 & 28.65 & 57.40 & 48.67 \\
SFT & 39.02 & 35.36 & 59.45 & 50.26 \\
Our CodeDPO & \textbf{42.07} & \textbf{38.04} & \textbf{61.37} & \textbf{53.43} \\
Our CodeKTO & 40.85 & 35.98 & 59.65 & 50.13 \\
\midrule
DeepSeekCoder-6.7B-base & 47.60 & 39.60 & 70.20 & 56.60 \\
SFT & 56.09 & 46.95 & 70.18 & 56.88 \\
Our CodeDPO & \textbf{59.75} & \textbf{51.83} & \textbf{72.18} & \textbf{60.01} \\
Our CodeKTO & 54.88 & 49.39 & 71.93 & 58.65 \\
\bottomrule
\end{tabular}
\caption{More results on the comparison of preference optimization strategies (DPO vs. KTO vs. SFT).}
\label{tab:ablationdpoAPP}
\vspace{-10pt}
\end{table*}

\section{Comparison with CodeT in details}

To clarify, CodeT \cite{chencodet} evaluates ranking based on the product of the number of test cases a piece of code passes and the number of codes that pass a given test case. While this approach provides a more refined evaluation than relying solely on the number of test cases passed, we observed in our experiments that it still falls short when compared to the fine-grained mechanism of our PageRank-based method. Specifically, CodeT's verification lacks the granularity offered by our PageRank-based method, particularly when sample size is relatively smaller. In our method, the sample size is set at 15, while CodeT generally uses a sample size of 100 (which is too large for us to follow due to resources).
We also give the detailed ablation result based on MagiCoder-S-DS-6.7B in Table \ref{tab:codetablation}, where ``Sort by $|C|*|T|$'' refers to the implementation of CodeT.

\begin{table*}[h]
\centering
\centering

\begin{tabular}{lcccc}
\toprule
Strategy & HumanEval & HumanEval+ & MBPP & MBPP+ \\
\midrule
MagiCoder-S-DS-6.7B & 73.17 & 68.29 & 76.72 & 66.67 \\
Filter with all tests & 76.22 & 68.90 & 77.44 & 67.17 \\
Sort by \# of passed tests & 78.66 & 70.73 & 79.19 & 68.67 \\
Sort by $|C|*|T|$ & 76.22 & 71.95 & 77.94 & 68.17 \\
Our CodeDPO & 83.54 & 76.22 & 80.70 & 70.93 \\
\bottomrule

\end{tabular}
\caption{Dataset Construction Ablations For MagiCoder-S-DS-6.7B}
\label{tab:codetablation}

\end{table*}

\section{CodeDPO Dataset Construction algorithm description}

In order to make it clear, we give a formal algorithm description of the CodeDPO construction pipeline in Algorithm \ref{alg:CodeDPO}.

The total pipeline includes four steps as shown in Algorithm \ref{alg:CodeDPO}:
\ding{182} Data Seed Construction with real-world source code: we extract concepts from code repositories and generate initial dataset;
\ding{183} Correctness Optimization with self-validation score: we generate initial code snippets and test cases, and update the self-validation score in loop;
\ding{184} Efficiency Optimization with execution time on credible tests: we record the execution time for each code snippet for ranking;
\ding{185} Model Optimization Training: we use the DPO loss for further training.

\begin{algorithm*}[h]
\footnotesize
\caption{CodeDPO Dataset Construction Pipeline}
\label{alg:CodeDPO}
\begin{algorithmic}[1]
\Procedure{CodeDPO}{model, instruction, max\_iterations}
    
    \State \textbf{Seed Construction:}
    \State Extract key programming concepts from source code repositories
    \State Generate code generation prompts and corresponding test cases
    \State Generate initial dataset $(instruction, solutions, test cases)$
    
    \State \textbf{Initialization:}
    \State Generate initial code snippets $C = \{c_1, c_2, ..., c_n\}$ from the instruction
    \State Generate test cases $T = \{t_1, t_2, ..., t_m\}$ corresponding to the instruction
    \State Initialize self-validation scores for code snippets and test cases: $\text{Score}(c_i) \gets 1$, $\text{Score}(t_j) \gets 1$
    \State Set damping factor $d \gets 0.85$
    \State $i \gets 0$
    
    \State \textbf{Self-Validation Loop:}
    \While{$i < \text{max\_iterations}$}
        \For{each $c_i \in C$}
            \State Execute $c_i$ on test cases $T$
            \For{each $t_j \in T$}
                \If{$c_i$ passes $t_j$}
                    \State Update $\text{Score}(c_i)$ using Equation (1)
                    \State Update $\text{Score}(t_j)$ using Equation (2)
                    \State \textbf{Execution Time Optimization:} 
                    \State Record execution time for $c_i$
                    \If{$c_i$ fails $t_j$}
                        \State Set execution time to max penalty to penalize $c_i$
                    \EndIf
                \EndIf
            \EndFor
        \EndFor
        \State $i \gets i + 1$
    \EndWhile
    
    \State \textbf{Final Dataset Collection:}
    \State \textbf{Correctness Optimization:}
    \State Select top-ranked code $c_{\text{top}}$ and low-ranked code $c_{\text{low}}$ for each instruction
    \State Store as dataset entries $(instruction, c_{\text{top}}, c_{\text{low}})$
    
    \State \textbf{Execution Time Optimization:}
    \State Select fastest code $c_{\text{fast}}$ and slowest code $c_{\text{slow}}$ for each instruction
    \State Store as dataset entries $(instruction, c_{\text{fast}}, c_{\text{slow}})$
    
    \State \textbf{return} final dataset entries
\EndProcedure
\end{algorithmic}
\end{algorithm*}

\section{LLM Prompts for CodeDPO dataset construction}

We use the following prompts for dataset seed construction and self-validation. 
During dataset construction, we first use code snippets from a randomly selected subset of \textbf{the Stack v1} dataset as input and prompt the LLM to generate the concept (LLM Prompt 1). Based on the concept, we then prompt the LLM to generate the task description (LLM Prompt 2).
For the validation process, we directly prompt the LLM with the task description to generate code solutions. Additionally, we prompt the LLM to generate only assertion statements as test cases (LLM Prompt 3). 
Since our chosen generation LLM is efficient and cost-effective, the entire process of data generation and construction takes around 40 hours on a server with 32 CPUs.

\begin{tcolorbox}[colframe=black!75!white, colback=white!95!black, boxrule=0.75mm, arc=5mm, outer arc=5mm, title=LLM Prompt \textbf{1} for Concept Generation]

Extract key programming concepts from a given code snippet collected from the open source repositories. Present the concepts as a comma separated list.

\begin{verbatim}
{Few-shot Examples}
\end{verbatim}

\#\# Example 2

\#\#\# Snippet

\begin{verbatim}
{Input Code}
\end{verbatim}

\#\#\# Concepts

\texttt{\textcolor{blue}{\{need to generate\}}}

\end{tcolorbox}

\begin{tcolorbox}[colframe=black!75!white, colback=white!95!black, boxrule=0.75mm, arc=5mm, outer arc=5mm, title=LLM Prompt \textbf{2} for Task Prompt Generation]

Create a set of independent code instructions that are original, different, diverse, and high-quality, where the properties control an instruction's category, language, concepts, and difficulty.

\begin{verbatim}
{Few-shot Examples}
\end{verbatim}

\#\# Example 2

\#\#\# Property
\begin{verbatim}
{Input Concept}
\end{verbatim}

\#\#\# Instruction

\texttt{\textcolor{blue}{\{need to generate\}}} 

\end{tcolorbox}

\begin{tcolorbox}[colframe=black!75!white, colback=white!95!black, boxrule=0.75mm, arc=5mm, outer arc=5mm, title=LLM Prompt \textbf{3} for Test Case Generation]

Generate only assertion statements based on the following description. Do not generate any other code:

\begin{verbatim}
{Instruction}
\end{verbatim}

Generated Assertions:

\texttt{assert \textcolor{blue}{\{need to generate\}}} 

\end{tcolorbox}




\section{\rev{Experiments on Challenging Code Efficiency Tasks}}

\rev{To evaluate code efficiency comprehensively, additional experiments are conducted on \textbf{EffiBench} \citep{huang2024effibench}. Since the absolute values of the results may vary depending on the specific execution environment, the analysis focuses on the relative improvements achieved by CodeDPO.}
\begin{table*}[h!]
    \centering
    \rev{
    \begin{tabular}{lcc}
        \toprule
        \textbf{Model} & \textbf{Total Execution Time} & \textbf{Normalized Execution Time} \\
        \midrule
        MagiCoder-S-DS-6.7B & 0.29 & 2.37 \\
        After CodeDPO & 0.21 & 1.58 \\
        \bottomrule
    \end{tabular}
    }
    \vspace{0.5cm}
    
    \rev{
    \begin{tabular}{lcc}
        \toprule
        \textbf{Model} & \textbf{Total Max Memory Usage} & \textbf{Normalized Max Memory Usage} \\
        \midrule
        MagiCoder-S-DS-6.7B & 24.71 & 1 \\
        After CodeDPO & 23.48 & 1 \\
        \bottomrule
    \end{tabular}
    \vspace{0.5cm}
    }

    \rev{
    \begin{tabular}{lcc}
        \toprule
        \textbf{Model} & \textbf{Total Memory Usage} & \textbf{Normalized Memory Usage} \\
        \midrule
        MagiCoder-S-DS-6.7B & 4.57 & 2.36 \\
        After CodeDPO & 3.90 & 1.93 \\
        \bottomrule
    \end{tabular}
    }
    \caption{\rev{Performance comparison on EffiBench for execution time and memory usage.}}
    \label{tab:effibench_results_combined}
\end{table*}

\rev{The results indicate that CodeDPO significantly reduces execution time and memory usage, both in absolute terms and after normalization, while maintaining comparable maximum memory usage. These improvements highlight the effectiveness of CodeDPO in optimizing code for both computational efficiency and resource usage, ensuring applicability to environments where performance and memory constraints are critical.}

\section{\rev{Execution Time for Code Efficiency Experiments}}

\rev{We present the average execution time (in seconds) for experiments conducted with the Phi-2-2.7B model. It is important to note that execution times may vary due to differences in computational resources and runtime conditions. To ensure the reliability of our measurements, repeated experiments are conducted in a stable environment, and the averaged statistics are reported below:}

\begin{table*}[h!]
    \centering
    \rev{
    \begin{tabular}{c|c|c|c}
        \hline
        \textbf{Benchmark} & \textbf{Before CodeDPO (s)} & \textbf{After CodeDPO (s)} & \textbf{Average Speedup} \\
        \hline
        HumanEval+ & 0.250 & 0.172 & 1.45x \\
        MBPP+      & 0.189 & 0.137 & 1.38x \\
        \hline
    \end{tabular}
    }
    \caption{\rev{Average execution time and speedup with CodeDPO.}}
    \label{tab:execution_time}
\end{table*}

\rev{These results demonstrate the consistent improvements in execution efficiency achieved through CodeDPO, highlighting its practical benefits in reducing runtime.}

\section{\rev{Ablation on Sample Number for code and test generation}}

\rev{The choice of the sample number and temperature, as described in Section \ref{sec:trainingsetting}, is guided by practical considerations to balance the diversity of sampled code solutions and test cases. These parameters are selected based on empirical observations and insights from prior work on data generation. To further investigate this, we conduct a series of ablation studies to evaluate the impact of varying sample numbers. Specifically, we tested sample numbers of 5, 15, and 50, with the experimental setup aligned with the design in Section \ref{sec:experimentAlg}.}
\rev{Table \ref{tab:spearman_corrandperformance_comparison} presents the Spearman correlation between the self-validation score and the actual code accuracy on the HumanEval dataset, and then shows the performance of the Phi-2-2.7B model for varying sample numbers, evaluated on both the HumanEval and HumanEval+ benchmarks. Similar trends are observed for other models. 
The results suggest that using \texttt{sample\_num=15} achieves a favorable trade-off between diversity and computational feasibility. While larger sample numbers provide marginal gains, they come with increased computational costs.}

\begin{table*}[h!]
    \centering
    \rev{
    \begin{tabular}{c|c|c|c|c}
        \hline
        \textbf{Sample Number ($n$)} & \textbf{Spearman Correlation} & & \textbf{HumanEval (\%)} & \textbf{HumanEval+ (\%)} \\
        \hline
        5  & 0.7425 & & 54.88 & 49.39 \\
        15 & 0.8598 & & 57.32 & 51.83 \\
        50 & 0.8613 & & 57.90 & 51.83 \\
        \hline
    \end{tabular}
    }
    \caption{\rev{Spearman correlation and performance of Phi-2-2.7B for different sample numbers.}}
    \label{tab:spearman_corrandperformance_comparison}
\end{table*}

\section{\rev{Discussion}}

\subsection{\rev{Comparison of Dataset Statistics}}

\rev{Since some baselines have not released their datasets, we rely on statistics reported in their respective papers for comparison. Below is a summary of dataset sizes and the number of unique questions, as both metrics are important—greater diversity in unique questions generally leads to higher dataset quality.}

\begin{table}[h!]
    \centering
    \rev{
    \resizebox{\linewidth}{!}{
    \begin{tabular}{lcc}
        \toprule
        \textbf{Method} & \textbf{Total Samples} & \textbf{Unique Questions} \\
        \midrule
        CodeDPO & 114k & 114k \\
        PLUM & Up to 120k & Up to 1,500 \\
        Code-Optimise & $\sim$100k (extended in our reproduction) & Up to 384 \\
        \bottomrule
    \end{tabular}
    }
    }
    \caption{\rev{Comparison of dataset sizes and unique questions across methods.}}
    \label{tab:dataset_sizes}
\end{table}

\rev{For SFT datasets, OSS-Instruct often combines multiple data sources. For example, models like MagiCoder-S-DS-6.7B and MagiCoder-S-CL-7B are trained using:}

\begin{table}[h!]
    \centering
    \rev{
    \begin{tabular}{lc}
        \toprule
        \textbf{SFT Dataset} & \textbf{Samples} \\
        \midrule
        Magicoder-OSS-Instruct & $\sim$75k \\
        Magicoder-Evol-Instruct & $\sim$110k \\
        Combined & Up to 185k \\
        \bottomrule
    \end{tabular}
    }
    \caption{\rev{Supervised fine-tuning dataset statistics.}}
    \label{tab:sft_datasets}
\end{table}

\rev{Based on comparisons with other related works, the dataset sizes of CodeDPO appear to be of the same order of magnitude. 
CodeDPO provides a significantly higher diversity in unique questions compared to baselines like PLUM and Code-Optimise, which heavily reuse prompts and have limited diversity despite similar sample sizes. This diversity ensures a more robust preference optimization process, which is a key advantage over existing approaches. 
}

\subsection{\rev{Implementation of the Self-Validation Scores}}

\subsubsection{\rev{Python Implementation of the Self-Validation Scores}}
\label{sec:pythonimp}
\begin{figure*}[t]
\centering
  \includegraphics[width=\linewidth]{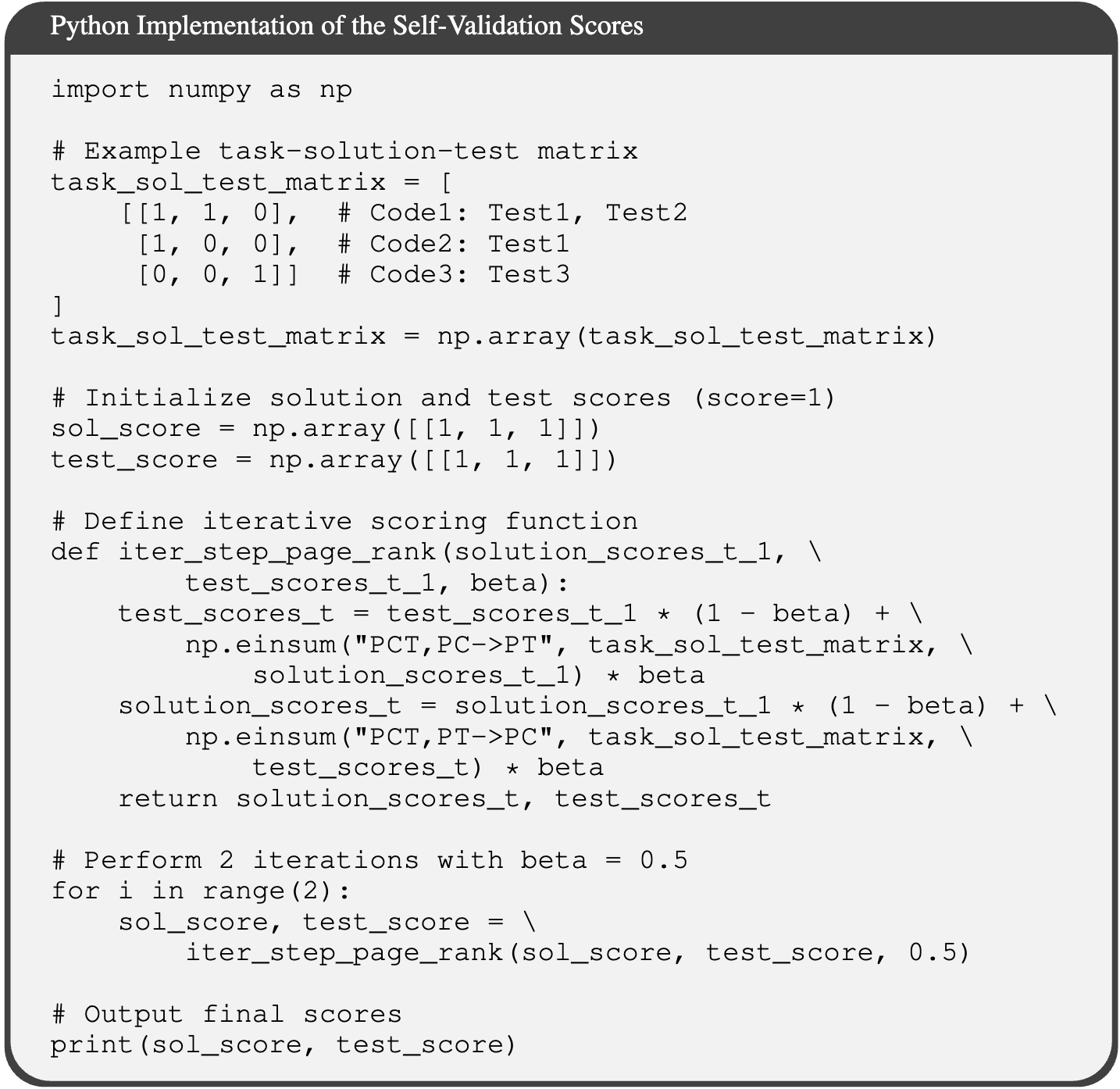}  
\caption{Python Implementation of the Self-Validation Scores in Figure \ref{fig:method}.}
\label{fig:pythonimp}
\end{figure*}

\rev{To enhance the understanding of the proposed algorithm, we provide a Python implementation illustrating the calculation process for the case in Figure \ref{fig:pythonimp} (specifically, Step 2 in the figure). The code demonstrates the iterative calculation of self-validation scores using a simplified example.}

\subsubsection{\rev{Handling Weak Test Cases}}
\rev{
Our designed algorithm is robust.
The self-validation scores can reflect the confidence of each code solutions and test cases through the iterative process. Notably, even in the presence of weak test cases (such as \texttt{assert True}), our method handles them robustly. 
We have carefully considered the impact of weak test cases in our design.
We address this issue from two perspectives:
\ding{182} \textbf{Natural Suppression of Weak Test Cases in Ranking:}
Weak test cases are those that almost all code solutions pass. While they contribute to the overall scores of all code solutions, they do not affect the relative differences between code solutions in the ranking process. Since the ranking is based on score differences, weak test cases naturally have minimal impact on the ranking outcomes.
\ding{183} \textbf{Filtering Identical or Close Scores:}
Weak test cases can lead to highly similar scores for multiple code solutions after repeated score updates, diminishing the ability to differentiate between them. To address this, as described in Section 3.4, we implement a filtering mechanism that excludes samples with identical or near-identical ranking scores. This ensures that the influence of weak test cases is mitigated in the final dataset.
}

\rev{
For example, assume we have 15 code solutions and 15 test cases generated by the model.
\ding{182} If a weak test case, such as assert True, is passed by all 15 code samples, its score during each update step (as computed by Equation 1) will contribute equally to the scores of all code solutions. As a result, it does not alter the relative ranking of the code solutions.
\ding{183} If all 15 test cases are similarly weak, the scores of the code solutions will converge to identical or near-identical values after several updates. To mitigate this, we apply a post-processing step (Section 3.4) to filter out such cases, ensuring the integrity of the final rankings.
By addressing weak test cases through these mechanisms, our algorithm achieves robustness and maintains the reliability of its outputs, even in challenging scenarios.
}

\end{document}